\documentclass[pra,twocolumn,superscriptaddress,showpacs,floatfix]{revtex4-1}
\usepackage{epsfig,graphicx,amsmath,amsfonts,amssymb}

\begin{document}
\title{Short-interaction quantum measurement through an incoherent mediator}

\author{J. Casanova}
\affiliation{Departamento de Qu\'{\i}mica F\'{\i}sica, Universidad del Pa\'{\i}s Vasco-Euskal Herriko Unibertsitatea, Apdo. 644, 48080 Bilbao, Spain}

\author{G. Romero}
\affiliation{Departamento de Qu\'{\i}mica F\'{\i}sica, Universidad del Pa\'{\i}s Vasco-Euskal Herriko Unibertsitatea, Apdo. 644, 48080 Bilbao, Spain}

\author{I. Lizuain}
\affiliation{Departamento de Qu\'{\i}mica F\'{\i}sica, Universidad del Pa\'{\i}s Vasco-Euskal Herriko Unibertsitatea, Apdo. 644, 48080 Bilbao, Spain}

\author{J. C. Retamal}
\affiliation{Departamento de
F\'{\i}sica, Universidad de Santiago de Chile, USACH, Casilla 307, Santiago 2, Chile}

\author{C. F. Roos}
\affiliation{Institut f\"ur Quantenoptik und Quanteninformation,
\"Osterreichische Akademie der Wissenschaften, Otto-Hittmair-Platz 1, A-6020 Innsbruck, Austria}
\affiliation{Institut f\"ur Experimentalphysik, Universit\"at Innsbruck, Technikerstr.~25, A-6020 Innsbruck, Austria}

\author{J. G. Muga}
\affiliation{Departamento de Qu\'{\i}mica F\'{\i}sica, Universidad del Pa\'{\i}s Vasco-Euskal Herriko Unibertsitatea, Apdo. 644, 48080 Bilbao, Spain}

\author{E. Solano}
\affiliation{Departamento de Qu\'{\i}mica F\'{\i}sica, Universidad del Pa\'{\i}s Vasco-Euskal Herriko Unibertsitatea, Apdo. 644, 48080 Bilbao, Spain}
\affiliation{IKERBASQUE, Basque Foundation for Science, Alameda Urquijo 36, 48011 Bilbao, Spain}

\date{\today}

\pacs{03.65.Ta, 42.50.Ex, 42.50.Dv}

\begin{abstract}

We propose a method of indirect measurements where a probe is able to read, in short interaction times, the quantum state of a remote  system through an incoherent third party, hereafter called {\it mediator}. Probe and system can interact briefly with the mediator in an incoherent state but not directly among themselves and, nevertheless, the transfer of quantum information can be achieved with robustness. We exemplify our measurement scheme with a paradigmatic example of this tripartite problem: qubit-oscillator-qubit, and discuss different physical scenarios pointing out the associated advantages and limitations.

\end{abstract}
\maketitle

\section{Introduction}

The transfer and extraction of information from quantum systems is an essential building block of quantum measurement theory~\cite{Zurek83}. This topic plays a key role in modern applications, like quantum computation and quantum information~\cite{nielsen}, quantum teleportation~\cite{bennett93}, and quantum cryptography~\cite{gisin02}. These applications have been driven by novel concepts in quantum information theory, as well as by successful measurement techniques achieved in different physical systems. Among them, we can mention electron-shelving qubit readout in trapped ions~\cite{leibfried03}, the quantum nondemolition measurement of field photon numbers and the measurement of Rydberg atom levels in microwave cavity QED~\cite{haroche06}. Recently, significant advances in the technology of quantum circuits~\cite{girvin08} have allowed efficient qubit readout techniques~\cite{wallraff05,serban08,mallet09}. From a theoretical point of view, there has been also recent interest in novel quantum measurement techniques, some of them aiming at improving experimental reach. We could mention weak measurements~\cite{Aharonov} in photonics~\cite{Lundeen} and circuit QED~\cite{Katz}, and short-interaction time measurements in the context of  trapped ions~\cite{Solano1,Gerritsma09,Zaehringer}, cavity QED~\cite{Solano2,Solano3}, and circuit QED~\cite{Serban1,Serban2}.

When a direct interaction of a quantum system with a probe is available, the system information can be transferred to the probe degrees of freedom. In this case, exchange of quantum information requires system-probe entanglement, but it does not warrant full reconstruction of the system quantum state or even the proper encoding of a searched system observable. In other cases we have to rely on indirect measurements, where a third party system mediates the communication between the system and the probe. In this sense, we are interested in the following question: is it possible that the probe measures fully the quantum state of a remote system in short interaction times if only their independent coupling to an intermediate incoherent mediator is allowed? In this work, we answer positively to this question and discuss the physical frame behind it. In Sec. II, we introudce our model, while Sec. III describes a method of measuring indirectly the quantum state of a remote system by encoding information in the short-time behavior. Both unitary and dissipative cases are studied. In Sec. IV, we propose a possible realistic implementation in trapped ion setups. Finally, we discuss in Sec. V possible errors via numerical analysis.

\section{The model}

\begin{figure}[b]
\vspace*{-0.7cm}
\centering
\includegraphics[width=0.5\linewidth]{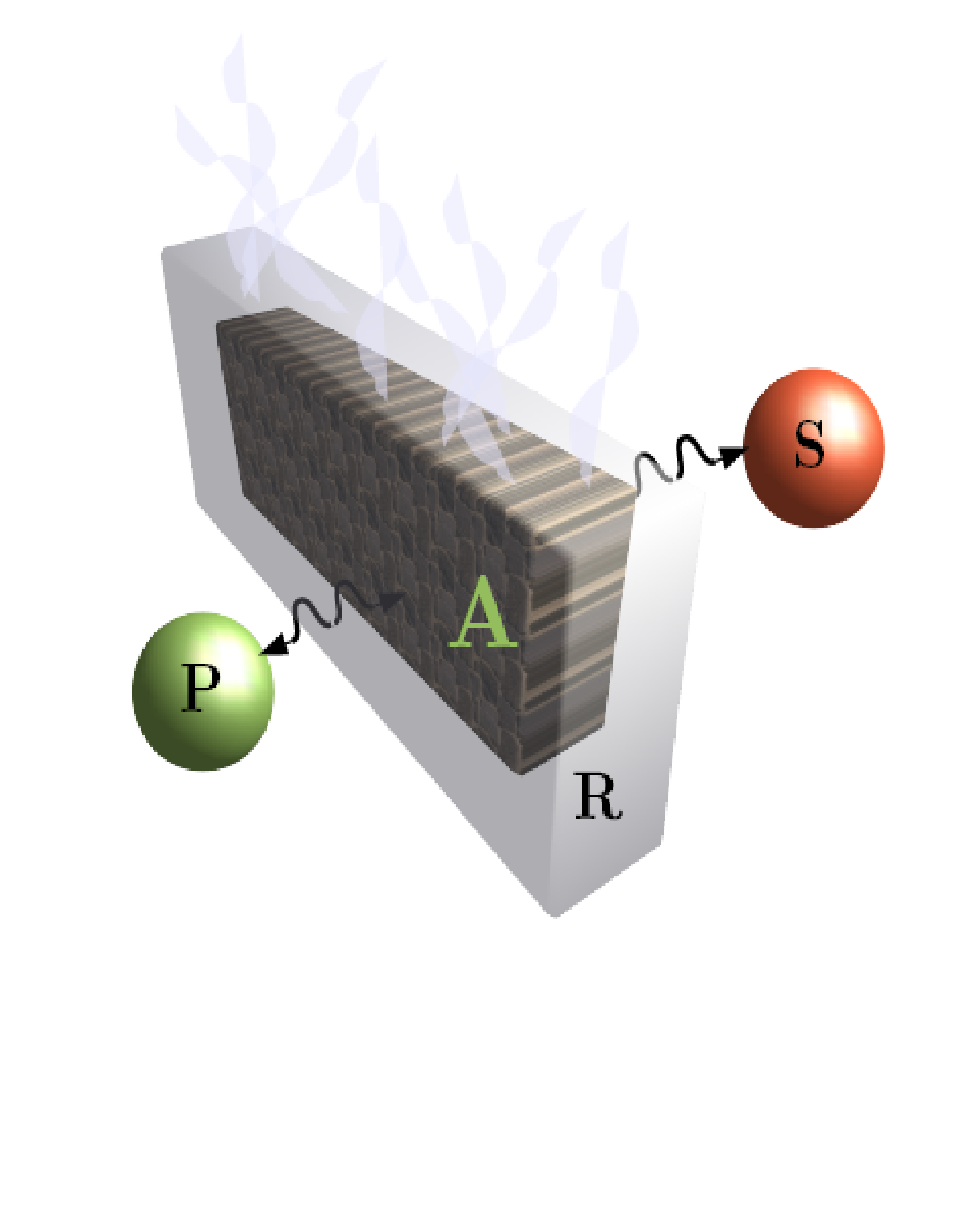}
\vspace*{-1.4cm}
\caption{(Color online) Sketch of the model. P, A, S, and R, are the probe, mediator, system, and reservoir, respectively.}
\label{fig1}
\end{figure}

We consider the following model: a probe denoted by $P$ will carry out the measurement of system $S$ through a mediator $A$. We may also consider the mediator in contact with a reservoir, as depicted in Fig.~\ref{fig1}. The associated total Hamiltonian reads

\begin{eqnarray}
\label{totalhamiltonian}
H = H_{P}+H_{S}+H_{A}+H_{P-A}+H_{S-A} ,
\end{eqnarray}
where $H_{P}$, $H_{S}$, and $H_{A}$, represent the free energies of the probe, system, and mediator, respectively, while the corresponding interactions are denoted by $H_{P-A}$ and $H_{S-A}$. No direct probe-system interaction is allowed, that is, the effective coupling betwen probe and system can only happen via higher-order terms of the unitary evolution. 

Let us consider a qubit(probe)-oscillator(mediator)-qubit(system) setup such that the initial tripartite state is decoupled, $\rho (0) = \rho_p(0) \otimes \rho_A(0) \otimes \rho_s(0)$. The unknown system state is described by the density operator
\begin{eqnarray}
\label{systemstate}
\rho_{s}(0) \! = \! \rho_{11} |1\rangle \langle1| + \rho_{12} |1\rangle \langle2| + \rho_{21} |2\rangle \langle1| + \rho_{22} | 2 \rangle \langle 2 | ,
\end{eqnarray}
which we want to measure with a probe in state $\rho_{p}(0) \! = \! \rho_{\rm gg} | {\rm g} \rangle \langle {\rm g} | + \rho_{\rm ge} | {\rm g} \rangle \langle {\rm e} | + \rho_{\rm eg} | {\rm e} \rangle \langle {\rm g} | + \rho_{\rm ee} | {\rm e} \rangle \langle {\rm e} |$. The free energy terms are $H_P = \frac{\hbar\omega_{p}}{2}\sigma^{z}_{p}$, $H_A = \hbar \omega_a a^{\dagger} a$, and $H_S = \frac{\hbar\omega_{s}}{2}\sigma^{z}_{s}$, while the interactions are of the Jaynes-Cummings type, $H_{P-A}=\hbar g_{p}(\sigma^{+}_{p} a + \sigma^{-}_{p} a^{\dag})$, $ H_{S-A}  =  \hbar g_{s}(\sigma^{+}_{s} a + \sigma^{-}_{s} a^{\dag})$.

\section{Short-interaction measurements}

\subsection{Unitary case}

As is the case in different experimental scenarios, we will consider the measurement of the probe observable $\theta_{p} = | {\rm e}\rangle \langle {\rm e}|$, $P_{\rm e}(\tau) = {\rm Tr} \left[ \rho (\tau) | {\rm e}\rangle \langle {\rm e}| \right]$, as a function of the dimensionless interaction time $\tau = g_p t$. From Eq.~(\ref{totalhamiltonian}), we derive
\begin{equation}
\frac{d P_{\rm e}(\tau)}{d \tau} = \frac{1}{i \hbar g_p} \langle[  | {\rm e}\rangle \langle {\rm e}| ,H_{P-A} ] \rangle
\end{equation}
and write
\begin{eqnarray}
\label{firstderivative}
\frac{d P_{\rm e}(\tau)}{d\tau} = \frac{1}{i} \left\langle \sigma_{p}^{+} a - \sigma_{p}^{-} a^{\dag} \right\rangle .
\end{eqnarray}
For $\tau = 0$, and choosing the initial probe state $\rho(0) = | +_{\phi} \rangle \langle +_{\phi} |$, where $| +_{\phi} \rangle = ( | {\rm g} \rangle + e^{i \phi} | {\rm e} \rangle ) / \sqrt{2}$, we get
\begin{eqnarray}
\left. \frac{d P_{\rm e}(\tau)}{d\tau} \right|_{\tau = 0} = \left\langle X_{\phi + \frac{\pi}{2}} \right\rangle .
\end{eqnarray}
This shows that any mediator quadrature, $X_{\phi} = (a^{\dagger} e^{i \phi} + a e^{-i \phi}) / 2$, can be measured via the first derivative of $P_{\rm e}(\tau)$ at vanishing interaction time~\cite{Solano1,Gerritsma09,Solano2,Solano3}. Due to the absence of direct probe-system coupling, it is clear from Eq.~(\ref{firstderivative}) that the probe is unable to record information about the system in a first-order expansion of $P_{\rm e}(\tau)$.

From Eq.~(\ref{totalhamiltonian}), we can derive further the second time derivative of $P_{\rm e}(\tau)$ at $\tau = 0$,
\begin{eqnarray}
\label{secondderivative}
\left. \frac{d^{2} P_{\rm e}(\tau)}{d\tau^{2}} \right|_{\tau = 0} = \left. \frac{1}{( i \hbar g_p)^2} \left< \bigg[ \big[  | {\rm e}\rangle \langle {\rm e}| , H_{P-A} \big] , H \bigg] \right> \right|_{\tau = 0} .
\end{eqnarray}
After using $[ | {\rm e}\rangle \langle {\rm e}| , H_{P-A} ] = \hbar g_{p} (\sigma_{p}^{+}a - \sigma_{p}a^{\dag} )$, the r.h.s. of Eq.~(\ref{secondderivative}) contains five remaining commutators associated with each term of $H$. The four nonvanishing elements, discarding $[ g_{p}(\sigma_{p}^{+}a - \sigma_{p}a^{\dag} ), \frac{\omega_{s}}{2}\sigma^{z}_{s}  ] = 0$, read
\begin{equation}
[ g_{p}(\sigma_{p}^{+} a - \sigma_{p}^{-} a^{\dag} ) ,  \frac{\omega_{p}}{2}\sigma^{z}_{p}] =  - g_{p}\omega_{p}(\sigma^{+}_{p}a +\sigma_{p}^{-} a^{\dag}  ) ,
\end{equation}
\begin{equation}
[ g_{p}(\sigma_{p}^{+}a - \sigma_{p}^{-} a^{\dag} ), \omega_a a^{\dag}a  ] = g_{p}\omega_a (\sigma^{+}_{p} a +\sigma_{p}^{-} a^{\dag}  ) ,
\end{equation}
\begin{equation}
[ g_{p}(\sigma_{p}^{+}a - \sigma_{p}^{-} a^{\dag} ), g_{p}(\sigma_{p}^{+} a + \sigma_{p}^{-} a^{\dag} )  ] = 2 g_{p}^{2}(\sigma_{p}^{+} \sigma_{p}^{-} + \sigma_{p}^{z}\hat{n}) ,
\end{equation}
\begin{equation}
g_{p} g_{s} [ (\sigma_{p}^{+}a - \sigma_{p}^{-} a^{\dag} ), (\sigma_{s}^{+} a + \sigma_{s}^{-} a^{\dag} )  ] \! = \! g_{p}g_{s} (\sigma_{p}^{+}\sigma_{s}^{-} + \sigma_{p}^{-} \sigma_{s}^{+}) . \vspace*{0.1cm}
\end{equation}
These terms are of second order in an expansion of $P_{\rm e}(\tau)$. We observe that the first two terms interfere destructively and cancel each other in resonance, $\omega_p = \omega_a$. The third term will be mediator independent when $\langle \sigma^z_p \rangle = 0$, that is when $\rho_{\rm gg} = \rho_{\rm ee}$. The fourth term is the only one that could map, in principle, system information onto probe degrees of freedom. Specifically, it associates system coherences with probe coherences, given that $\rho_{\rm ge} = \rho^*_{\rm eg} \neq 0$. It is clear, then, that a good choice of an initial probe state is $\rho_{p}(0) = |+_{\phi} \rangle \langle +_{\phi} |$. Note that in most quantum optical setups, where the probe is a two-level atom, the fidelity in the generation of this state is very high. Nevertheless, we will estimate possible errors in the last section of this work. Finally, it can be shown that
\begin{eqnarray}
\label{result}
\left. \frac{d^{2} P_{\rm e}(\tau)}{d\tau^{2}} \right \arrowvert_{\tau = 0}  \!\!\! = -1 + \frac{\delta}{g_{p}} \langle X_{\phi}\rangle - \frac{g_{s}}{2g_{p}}( \rho_{\rm 12} e^{i\phi} + \rho_{\rm 21} e^{-i\phi} ) , \nonumber \\
\end{eqnarray}
where $\delta = \omega_{p} - \omega_a$ is the detuning between the probe and the mediator. If the field is incoherent, $\rho=\sum P_{n}\arrowvert n\rangle\langle n\arrowvert$, e.g. a thermal state with $\langle X_{\phi}\rangle = 0$, or under the resonant condition $\delta = 0$, we are left with
\begin{equation}
\label{result2}
\left.\frac{d^{2}P_{\rm e}(\tau)}{d\tau^{2}}\right \arrowvert_{\tau = 0} = -1 - \frac{g_{s}}{2g_{p}}( \rho_{\rm 12} e^{i\phi} + \rho_{\rm 21} e^{-i\phi} ) .
\end{equation}
This result shows that, though the probe and the system are only indirectly connected via an mediator, incoherent or not, it is possible to encode the coherences of the system in the short-time behavior of the second derivative of the probe's $ P_{\rm e}(\tau)$. A proper choice of phase $\phi$ in the probe state $|+_{\phi} \rangle$ will allow us to read the real and imaginary parts of the system coherences. We remark that an incoherent mediator does not prevent the system to implement a fast transfer of information to the probe, as can be seen in Eqs.~(\ref{result}) and (\ref{result2}).

From previous considerations, we cannot gain information about the system populations due to the structure of the interactions $H_{P-A}$ and $H_{A-S}$. Note that these interaction terms only couple, in first-order, off-diagonal elements of each qubit and the mediator degrees of freedom. Given that the expression in Eq.~(\ref{result2}) is related to a qubit-mediator-qubit second-order process, the system coherences can be, in principle, mapped onto the second derivative of the probe population. Nevertheless, we can measure the system populations $\rho_{\rm 11}$ and $\rho_{\rm 22}$ provided we move from the quasiresonant regime to the off-resonant limit, where the qubits are dispersively coupled to the mediator. For the sake of simplicity, we will consider here that both qubits have the same transition frequency, $\omega_p = \omega_s$, and are coupled to the mediator with the same coupling strength $g = g_p = g_s$. In this case, we can write a second-order effective Hamiltonian from Eq.~(\ref{totalhamiltonian}) in the dispersive regime, $\delta \gg g \sqrt{\langle n \rangle}$, as
\begin{eqnarray}
\label{dispersivehamiltonian}
H_{\rm eff} \! = \! \frac{\hbar g^2}{\delta} \big\lbrack (\sigma_p^z + \sigma_s^z ) a^{\dagger} a + ( \sigma_p^{+} + \sigma_s^{+} ) ( \sigma_p^{-} + \sigma_s^{-} ) \big\rbrack .
\end{eqnarray}
Remark that in this second-order Hamiltonian, as expected, direct coupling terms between the qubits are effectively generated.
In this case, following similar steps as the ones used to obtain Eq.~(\ref{result2}), we can derive
\begin{equation}
{\label{populations}}
\left.\frac{d^{2}P_{\rm e}(\tau)}{d\tau^{2}}\right\arrowvert_{\tau = 0} = \frac{g^{2}}{\delta^{2}} \ (\rho_{\rm 22} - \rho_{\rm 11}) .
\end{equation}
It is worth mentioning that the result of Eq.~(\ref{populations}), similar to Eq.~(\ref{result}), can also work with any particular mediator state. In fact, this insensitivity could be used to test whether the couplings are quasiresonant or dispersive. From Eqs.~(\ref{result2}) and (\ref{populations}), we can then recover the full density matrix describing the system, see Eq.~(\ref{systemstate}), which is encoded in the short-time behavior of the probe population. In the last section of this article, we will study these results with analytical and numerical tools.

\subsection{Dissipative case}

We will study now tougher scenarios where the previous results may fail to work. Let us consider then a single-mode quantum harmonic oscillator as the mediator with a thermal bath acting on it~\cite{Schweiger}. The associated master equation in the Markov approximation reads
\begin{eqnarray}
\dot{\rho} = \frac{1}{i \hbar} [ H , \rho ] + \gamma \frac{\bar{n}_b}{2}[2 a^{\dag} \! \rho a -  a a^{\dag} \! \rho - \rho a a^{\dag}] \nonumber \\  + \frac{\gamma}{2}(\bar{n}_b+1)[2 a \rho  a^{\dag} - a^{\dag}  a \rho - \rho  a^{\dag}  a] ,
\end{eqnarray}
where $\bar{n}_b$ is the mean number of bath thermal excitations, and we will denote as $\bar{n}_a$ the mean number of thermal excitations associated with the initial mediator state. We will consider the effects of the reservoir on our proposed readout of the system coherences. Surprisingly, we find that the presence of the reservoir does not affect the second derivative of $P_{\rm e}(\tau)$, preserving the results displayed in Eqs.~(\ref{result2}) and (\ref{populations}). It can be proven that the earliest bath contribution appears only in the third-order derivatives,
\begin{eqnarray}
\!\!\!\!\!\!\!\!\!\! \frac{d^3 P_{\rm e}(\tau)}{d\tau^3} \!\! & = & \!\!\! \frac{1}{(i \hbar)^3} {\rm Tr} \bigg\lbrace{ \! \rho(\tau) \! \bigg[ \big[ [ | e \rangle\langle e | , H_{P-A} ] , H \big] , H \bigg] \! \bigg\rbrace} \nonumber \\  & \ & + \frac{1}{(i \hbar)^2} {\rm Tr} \bigg\lbrace{ \! \mathcal{L}\rho(\tau) \big[ [ | e \rangle\langle e | , H_{P-A} ] , H \big] \! \bigg\rbrace} ,
\label{thirdderivative}
\end{eqnarray}
where the terms in the r.h.s. account for the unitary and nonunitary corrections to the second-order derivatives.
If the initial probe state is
$\alpha|{\rm g} \rangle + \beta | {\rm e}\rangle$, the correction to Eq.~(\ref{result}), for instance, reads
\begin{equation}
\frac{\gamma}{g}(|\alpha|^2-|\beta|^2)(\bar{n}_a-\bar{n}_b) .
\label{thirdderivative}
\end{equation}
It is noteworthy to mention that this contribution of the thermal reservoir becomes null whenever $| \alpha | = | \beta |$ or $\bar{n}_a = \bar{n}_b$.  The first condition can be easily satisfied when preparing the initial probe state, as in $|+_{\phi}\rangle$, though we will discuss it in more detail below. The second condition is related to the lack of initial thermalization between the mediator and the bath. The fact that the reservoir affects Eqs.~(\ref{result2}) and (\ref{populations}) only in higher orders is a signature of robustness of our proposed measurement scheme.

To make a final test of the introduced concepts, we will consider now the case of both qubits interacting with a bath, that is the mediator consists of a continuum of modes. This model can be described with the Hamiltonian
\begin{eqnarray}
\label{Eq18}
H =  \!\!\!\!\!\! && \frac{ \hbar \omega}{2} \sigma _{s}^{z}+\frac{ \hbar \omega}{2} \sigma _{p}^{z}  + \hbar \sum \limits_{k}  \omega _{k} b_{k}^{\dagger } b_{k}  \nonumber \\ && + \hbar \sum \limits_{k} g_{k} \big\lbrack (\sigma _{s}^{-} + \sigma_{p}^{-}) b_{k}^{\dagger} + (\sigma _{s}^{+} + \sigma _{p}^{+} ) b_{k} \big\rbrack
\end{eqnarray}

We consider, for the sake of simplicity, similar qubit transition
frequencies, $\omega = \omega_s = \omega_p$, and similar coupling between each mode and the qubits. This model contains some features of the unitary model described by Eq.~(\ref{totalhamiltonian}), only that now the single mode has been replaced by many. It should be clear that both qubits are not coupled directly but they will get effectively entangled via the mediator modes in higher orders of the evolution. However, it is also expected that these modes play the role of a dissipative bath, making difficult the exchange of quantum information between the qubits. As a further scope, we would want to remark that going beyond the lowest-order perturbative regime of Eq.~(\ref{Eq18}), a wealth of known effects appear in quantum information~\cite{Verstraete09}. After tracing the bath degrees of freedom and using the Markov approximation, we are left with the master equation~\cite{Schneider},
\begin{eqnarray}
\label{lastmasterequation}
\dot{\rho} =  \!\!\!\!\!\! && \frac{1}{i \hbar} [ H_0 , \rho ] + \frac{\varGamma}{2} \left[ 2 ( \sigma_{p}^{-} + \sigma_{s}^{-} ) \rho( \sigma_{p}^{+}  + \sigma_{s}^{+} ) \right. \nonumber \\ && \left. - (\sigma_{p}^{+} + \sigma_{s}^{+} ) (\sigma_{p}^{-} + \sigma_{s}^{-}) \rho \! - \! \rho(\sigma_{p}^{+} + \sigma_{s}^{+} ) ( \sigma_{p}^{-} + \sigma_{s}^{-} ) \right] , \nonumber \\
\end{eqnarray}
where $H_0 = \frac{ \hbar \omega}{2} \sigma _{s}^{z}+\frac{ \hbar \omega}{2} \sigma _{p}^{z} $. In this case, and with $\arrowvert+_{\phi}\rangle$ as the initial probe state, we can calculate
\begin{equation}
\label{bath}
\left.\frac{d^{2}P_{\rm e} (\tau)}{d \tau^{2}} \right\arrowvert_{\tau = 0} \!\!\! = \! \frac{1}{4} \! \left[ (\rho_{\rm 11} \! - \! \rho_{\rm 22}) \! + 2 (1 + \rho_{\rm 12} e^{i \phi} + \rho_{\rm 21} e^{-i \phi}) \right] ,
\end{equation}
where the dimensionless time is now $\tau = \varGamma t$.
We observe that the populations and coherences of the unknown system are revealed all together at short-interaction times. This result can be intuitively explained in the following manner. The reservoir model for both qubits assumes the presence of a continuum of modes with different frequencies, some of them closer to resonance and some dispersively coupled to the qubits. Equation~(\ref{bath}) reflects, in an averaged manner, the influence of both cases, and it can be considered as related to the results found in Eqs.~(\ref{result2}) and (\ref{populations}). In consequence, it is possible to read the quantum information of a system with a probe only by the fact of sharing a common reservoir, as expressed by Eqs.~(\ref{lastmasterequation}) and (\ref{bath}). Though this result seems to be strong, we must bear into account the influence of more realistic bath parameters. When $\varGamma$ is large, as expected for a real bath, a given discrete step of the dimensionless time $\tau =\varGamma t$ would require a reduction in the step of the lab interaction time~$t$. In consequence, a large thermal bath acting on both qubits would make more difficult for the probe to extract relevant system information.

\section{Experimental issues}

\begin{figure}[h]
\centering
\includegraphics[width=0.95\linewidth]{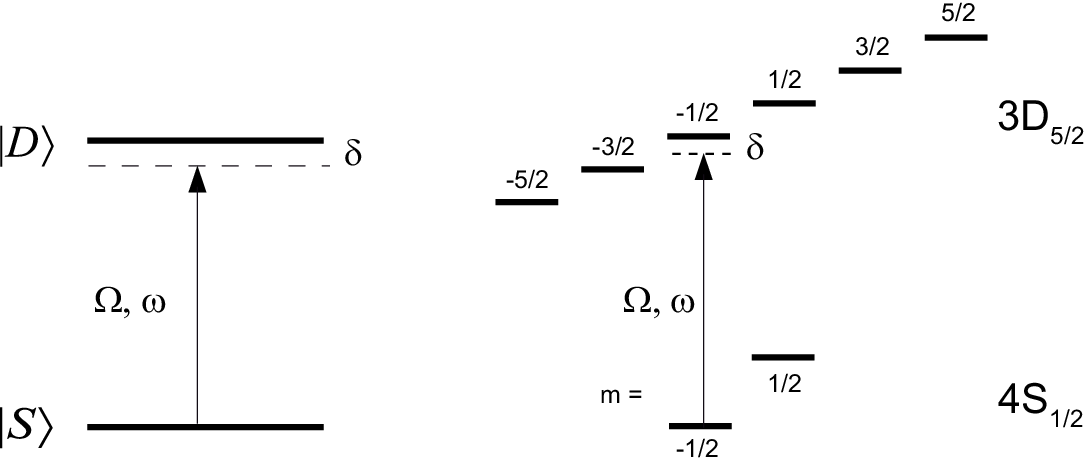}
\caption{Realization of a qubit in a single trapped $^{40}$Ca$^+$ ion. The qubit states $|S\rangle$, $|D\rangle$ are encoded in a pair of Zeeman states interacting with a laser pulse of frequency
$\omega$ and coupling strength $\Omega$.} \label{fig:niveles}
\end{figure}

The present ideas could be implemented and tested in different physical setups.  Due to their advanced precision in quantum measurements, we believe that circuit QED and trapped ion experiments are good candidates, and we give details for the latter. We consider two trapped and laser-manipulated ions coupled through the vibrational center-of-mass motion like, for example, $^{40}\rm{Ca}^{+}$ ions held in a linear ion trap~\cite{SchmidtKaler03,Benhelm08} with center-of-mass oscillation frequency $\nu$. In this case, each qubit can be realized by using the electronic level $|4S_{1/2}(m=-1/2)\rangle \equiv|S\rangle$ as the
ground state and meta-stable level $|3D_{5/2}(m=-1/2)\rangle \equiv|D\rangle$ as the
excited state~\cite{SchmidtKaler03}, as shown in Fig.~\ref{fig:niveles}, with a magnetic field lifting the degeneracy of the Zeeman levels. In order to implement our measurement protocol, we consider monochromatic laser pulses individually addressing a single ion of the ion crystal~\cite{Nagerl99} with a frequency $\omega$ close to the qubit transition frequency $\omega_0$. After applying an optical RWA, the Hamiltonian describing the ion-laser interaction reads
\begin{equation}
H^{I}_{\mathrm{int}}=\frac{\hbar\Omega}2 \left( |D\rangle \langle
S|e^{i\eta(a^{\dag }e^{i\nu t}+ a e^{-i\nu t})}e^{i(\varphi-\delta
t)}+ {\rm H.c.} \right) . \label{Hint}  \vspace*{0.1cm}
\end{equation}
Here, $\Omega$ is the coupling strength, $\delta=\omega-\omega_0$ the detuning, and $\eta=k (\hbar/2m
\nu)^{1/2}$ is the Lamb-Dicke parameter defined as the square root of the ratio between the
recoil energy and the ion center of mass quantum vibration energy.
Operators $a$, $a^{\dag }$ describe annihilation and creation of vibrational excitations of the center-of-mass mode.

In the case where the laser is tuned to the red sideband resonance, $\delta= -\nu$, the interaction (\ref{Hint}) turns into a Jaynes-Cummings Hamiltonian
\begin{equation}
{H}_R =\frac{i\eta \hbar \Omega}{2}(a^{\dag }|S\rangle \langle
D|e^{-i\varphi}- a |D\rangle \langle S|e^{i\varphi}),  \label{red}
\end{equation}
when the vibrational RWA is carried out and the ion is assumed to be in the Lamb-Dicke regime ($\eta\langle  \bar{n}\rangle \ll 1$). Here, the phase $\varphi$ corresponds to the relative
phase between the optical field and the atomic polarization. For $\varphi=0$, the Hamiltonians $H_{P-A}$ or $H_{S-A}$ of Eq.~(\ref{totalhamiltonian}) are realized. Measurement of the probe ion-qubit as required by (\ref{result2}) or (\ref{populations}) is efficiently carried out via electron shelving~\cite{leibfried03}. In an ion trap experiment, the coupling strength $g=\eta\Omega$ needs to be kept much smaller than the oscillation frequency $\nu$ in order to avoid off-resonant excitation of the ion on the carrier transition. For this reason, $P_{\rm e} (\tau)$ changes on a time scale that is slow compared with the time scale $\propto\nu$ on which off-resonant excitations occur. Measurement of first and second derivatives of $P_{\rm e} (\tau)$ have been already accomplished in the lab with high precision~\cite{Zaehringer,Gerritsma09}. They were implemented by measuring $P_{\rm e}(\tau)$ for different probe times and fitting the measured signal by a low-order polynomial. Measurement of $P_{\rm e} (\tau)$ for a fixed probe time $\tau$ entails preparing and detecting the ions a large number of times to keep the quantum projection noise at an acceptable level. Detecting, for example, $1\%$ changes of $P_{\rm e} (\tau)$ requires more than $10^4$ experimental repetitions which could be realised within a few minutes in current experiments.

\begin{figure}[t]
\centering
\includegraphics[width=0.85\linewidth]{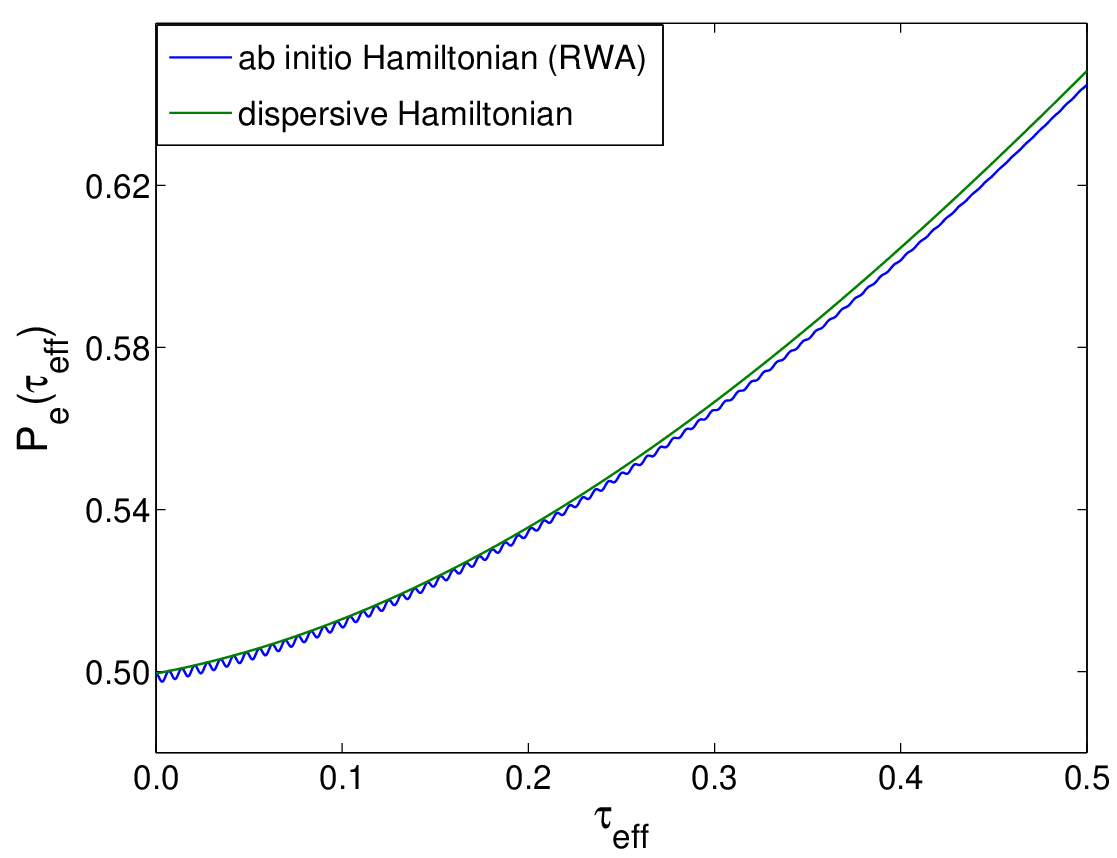}
\includegraphics[width=0.85\linewidth]{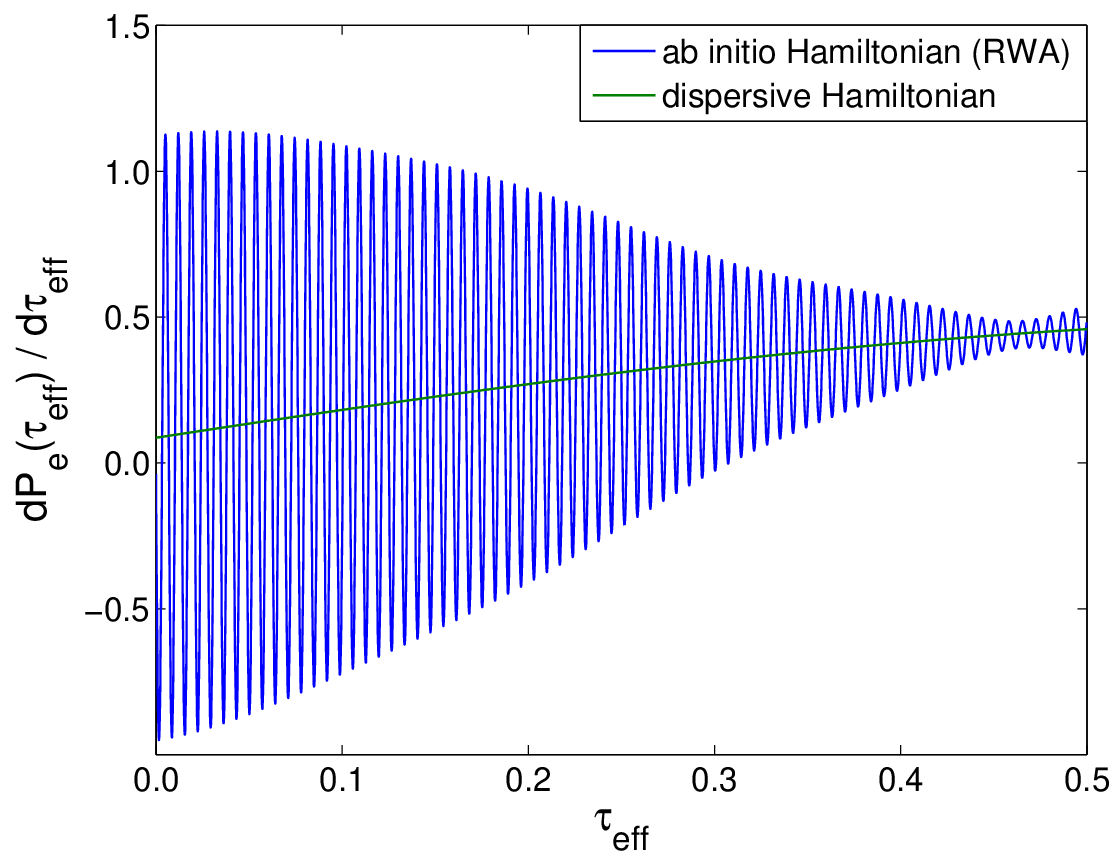}
\includegraphics[width=0.85\linewidth]{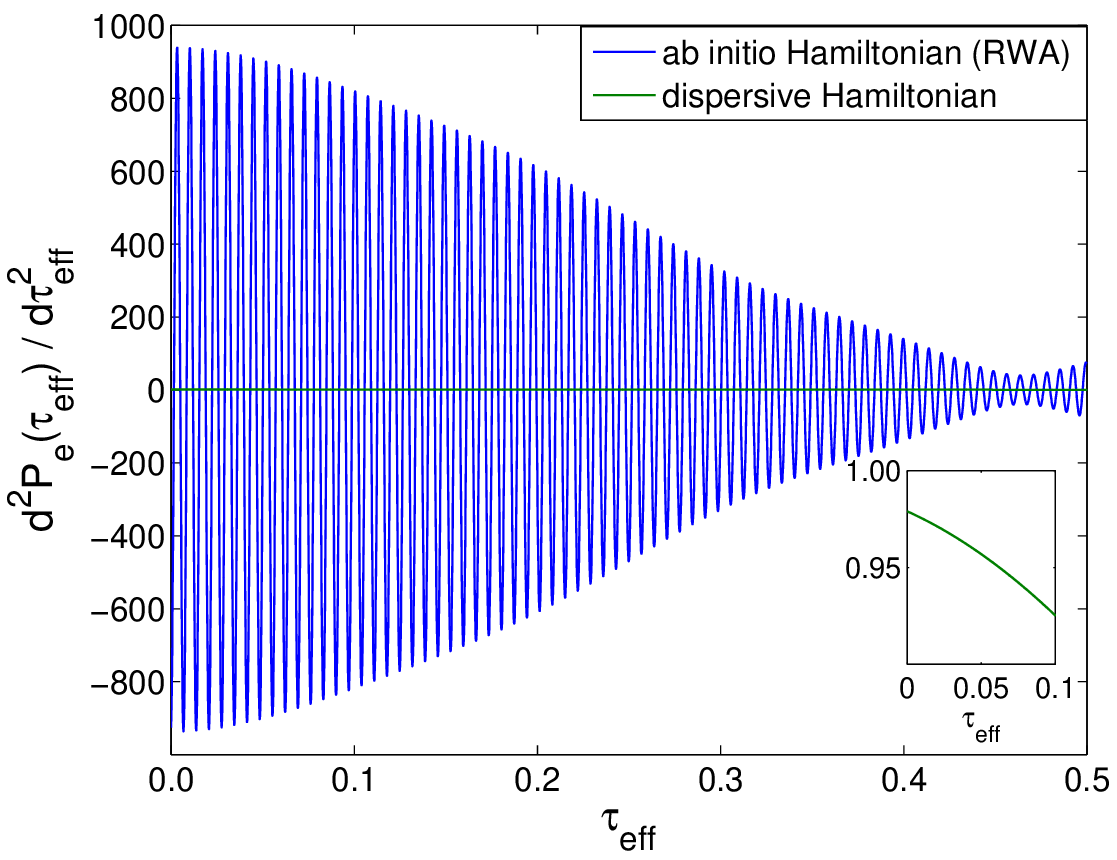}
\caption{(Color online) Plots of $P_{\rm e}(\tau_{\rm eff})$ and its derivatives as a function of the dimensionless effective time, $\tau_{\rm eff} = g_p^2 t / \delta$, for a dispersive case where the system state is $| \psi_s \rangle = 0.1 | 1 \rangle + e^{i \pi / 3} \sqrt{1 - 0.1^2} | 2 \rangle$, the probe state is $| \psi_p \rangle = ( | {\rm g} \rangle + | {\rm e} \rangle ) / \sqrt{2}$, the detuning is $\delta = 30g_p$, and the mediator is in a thermal state with $\bar{n}_a = 1$. The fast oscillating blue lines correspond to a calculation done with the ab initio Hamiltonian of Eq.~(\ref{totalhamiltonian}) and the green line to the dispersive Hamiltonian of Eq.~(\ref{dispersivehamiltonian}) .}
\label{fig3}
\end{figure}

\section{Discussion}

In this work, we deal with (infinitesimal) derivatives of an expansion of $P_{\rm e}(\tau)$ under some initial conditions and approximations. From this point of view, we should be careful when comparing our theoretical results with measured discrete derivatives in the lab. These considerations are not very different from the usual ones in similar models, but special attention should be given to derivatives. For example, Eq.~(\ref{result2}) is valid only if we are not allowed or do not want to observe the probe dynamics with a temporal resolution beyond the rotating-wave-approximation (RWA). In this case, the second derivative of Eq.~(\ref{result2}) predicts a correct measured value in the lab if the minimal discrete time interval of the experimental sampling is $\Delta \tau \geq g_p / \omega_p$. Otherwise, we would have to recalculate the model beyond the RWA to make the correct predictions. Similar arguments follow for the second derivative of Eq.~(\ref{populations}). Here, it is not only the RWA discreteness that should be requested, it is also the adiabaticity of the second-order effective Hamiltonian that imposes an even slower temporal step in the dispersive limit. In consequence, the results of Eqs.~(\ref{result}) and (\ref{populations}) cannot be compared when $\delta$ is large.  In this sense, we have carried out numerical simulations to study and test the dispersive case for a particular example. Both, the ab initio Hamiltonian of Eq.~(\ref{totalhamiltonian}) and the dispersive Hamiltonian of Eq.~(\ref{dispersivehamiltonian}) were used, under similar conditions, to test the predictions of our proposed protocol. For the ab initio model, the blue lines of Fig.~\ref{fig3} show the fast oscillatory evolution that cannot be observed in experiments due to limitations in the temporal resolution. For the effective model, that of Eq.~(\ref{dispersivehamiltonian}), the green lines show the expected measured curves and derivatives. Note that the inset of the last plot in Fig.~\ref{fig3} shows the measured value of the second derivative, $~0.975$, at $\tau_{\rm eff} = 0$.

\begin{figure}[t]
\centering
\includegraphics[width=0.75\linewidth]{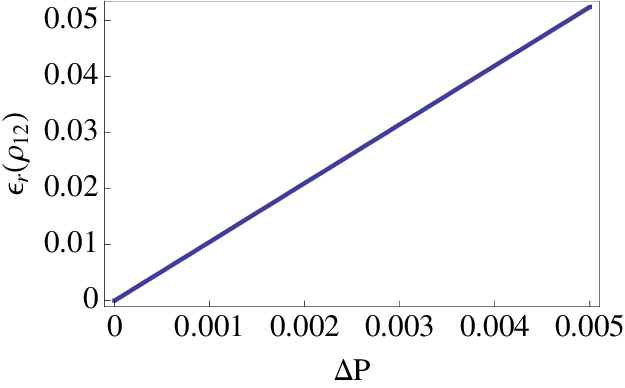}
\includegraphics[width=0.75\linewidth]{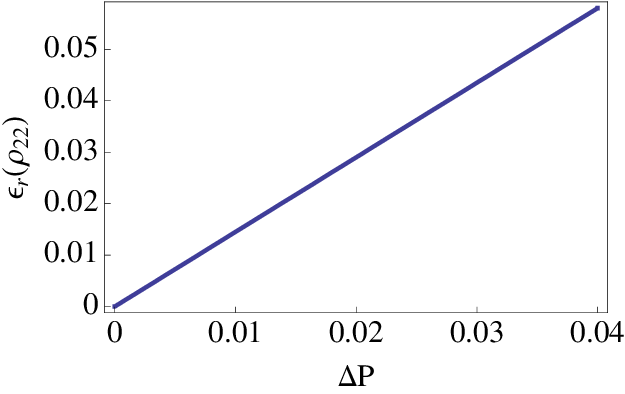}
\includegraphics[width=0.75\linewidth]{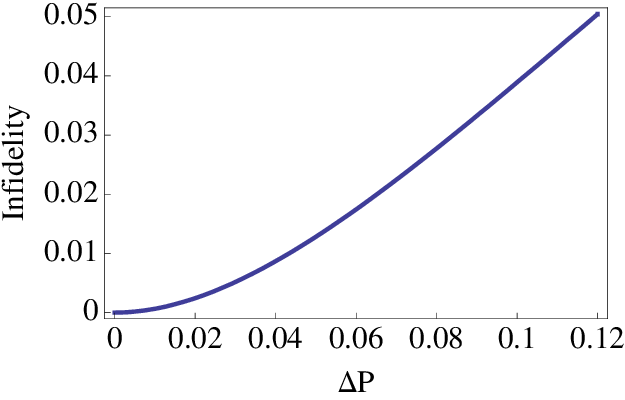}
\caption{(Color online) Plots of the relative erros in the measurement of the system coherence and population, $\epsilon_{\rho_{12}}$ and $\epsilon_{\rho_{22}}$, respectively, and the infidelity, $1-F$, of the measured density matrix as a function of the population inversion $\Delta {\rm P}$. The state of the system is $| \psi_s \rangle = 0.3 | 1 \rangle + \sqrt{1 - 0.3^2} | 2 \rangle$ and that of the probe is $| \psi_p \rangle = ( | {\rm g} \rangle + | {\rm e} \rangle ) / \sqrt{2}$ .}
\label{fig4}
\end{figure}

We have also tested the robustness of our proposed scheme when deriving the first key result in Eq.~(\ref{result2}). We have considered imprecisions in the required vanishing of the population inversion of the probe, $\Delta {\rm P} = \langle \sigma_p^z \rangle$ as displayed in Fig.~\ref{fig4}. In this figure, we have plotted the relative errors in the measurement of the system coherence and population, $\epsilon_{\rho_{12}} = ( \rho_{12} - \rho_{12}^{\rm exp} ) / \rho_{12}$ and  $\epsilon_{\rho_{22}} = ( \rho_{22} - \rho_{22}^{\rm exp} ) / \rho_{22}$, respectively, as a function of $\Delta {\rm P}$. Furthermore, we have plotted the infidelity of the measured state, $1-F$, also as a function of $\Delta {\rm P}$. Here, the fidelity $F$ was defined with the help of the Frobenius norm and inner product as $ F = {\rm Tr} [ \rho^{\rm exp }  \rho^{\dagger } ] /  \sqrt{ {\rm Tr} [ \rho^{\rm exp } \rho^{\rm exp \dagger} ] } \sqrt{ {\rm Tr} [ \rho \rho^{\dagger} ] }$. From the plots in Fig.~\ref{fig4}, as a key feature, one can observe that an imprecision of $0.01$ in the population inversion of the probe will lead to a fidelity much better than $99 \%$, assuring the protocol robustness under $\Delta {\rm P}$ fluctuations.

\section{Conclusions}

We have introduced a quantum measurement technique where a probe is able to read, fast and efficiently, the quantum information of a remote system under different severe conditions. Though we are confident of its wide applicability, we have discussed a possible implementation of it in trapped ions. Finally, we have discussed key features of the model and some fidelity issues.

\acknowledgments

J.C. acknowledges financial support from the Basque Government BF108.211, G.R. from Juan de la Cierva Program, and J.C.R. from Fondecyt 1070157. E.S. acknowledges funding from UPV-EHU GIU07/40, Ministerio de Ciencia e Innovaci\'on FIS2009-12773-C02-01, EuroSQIP and SOLID European projects.

\end{document}